\documentclass[journal]{IEEEtran}
\usepackage[utf8]{inputenc}
\usepackage[numbers]{natbib}
\usepackage{graphicx}
\usepackage{authblk}
\usepackage[pdftex,dvipsnames, table]{xcolor}
\usepackage[stable, bottom]{footmisc}
\usepackage{longtable,ltcaption}
\usepackage{setspace}
\usepackage{rotating}
\usepackage{eurosym}
\usepackage{lscape}

\usepackage{soul} 
\usepackage[margin=1in]{geometry}
\usepackage{comment}
\usepackage[normalem]{ulem}

\usepackage{CJKutf8}

\pdfoutput=1

\definecolor{lightgray}{gray}{0.9}
\let\oldtabular\tabular
\let\endoldtabular\endtabular
\renewenvironment{tabular}{\rowcolors{2}{white}{lightgray}\oldtabular}{\endoldtabular}

\let\oldlongtable\longtable
\let\endoldlongtable\endlongtable
\renewenvironment{longtable}{\rowcolors{2}{white}{lightgray}\oldlongtable} {
\endoldlongtable}

\newcounter{magicrownumbers}
\newcommand\rownumber{\stepcounter{magicrownumbers}\arabic{magicrownumbers}}

\begin{document}
	
\title{Cyber Security in the Age of COVID-19: A Timeline and Analysis of Cyber-Crime and Cyber-Attacks during the Pandemic}

\renewcommand\Affilfont{\fontsize{9}{8}\itshape}

\author[1]{Harjinder Singh Lallie}
\author[2]{Lynsay A. Shepherd}
\author[3]{Jason R. C. Nurse}
\author[4]{Arnau Erola}
\author[1]{Gregory Epiphaniou}
\author[1]{Carsten Maple}
\author[5]{Xavier Bellekens}

\affil[1]{WMG, University of Warwick, \{HL, CM, gregory.epiphaniou\}@warwick.ac.uk}
\affil[2]{School of Design and Informatics, Abertay University, lynsay.shepherd@abertay.ac.uk}
\affil[3]{School of Computing, University of Kent, j.r.c.nurse@kent.ac.uk}
\affil[4]{Department of Computer Science, University of Oxford, arnau.erola@cs.ox.ac.uk}
\affil[5]{Department of Electronic and Electrical Engineering, University of Strathclyde, xavier.bellekens@strath.ac.uk}

\date{}
\maketitle{}

	\begin{abstract}
		The COVID-19 pandemic was a remarkable unprecedented event which altered the lives of billions of citizens globally resulting in what became commonly referred to as the \textit{new-normal} in terms of societal norms and the way we live and work. Aside from the extraordinary impact on society and business as a whole, the pandemic generated a set of unique cyber-crime related circumstances which also affected society and business. The increased anxiety caused by the pandemic heightened the likelihood of cyber-attacks succeeding corresponding with an increase in the number and range of cyber-attacks.

This paper analyses the COVID-19 pandemic from a cyber-crime perspective and highlights the range of cyber-attacks experienced globally during the pandemic. Cyber-attacks are analysed and considered within the context of key global events to reveal the modus-operandi of cyber-attack campaigns. The analysis shows how following what appeared to be large gaps between the initial outbreak of the pandemic in China and the first COVID-19 related cyber-attack, attacks steadily became much more prevalent to the point that on some days, 3 or 4 unique cyber-attacks were being reported. The analysis proceeds to utilise the UK as a case study to demonstrate how cyber-criminals leveraged key events and governmental announcements to carefully craft and design cyber-crime campaigns. 

	\end{abstract}
	
	Coronavirus; COVID-19; cyber security; cyber-attack; cyber-crime; attack timeline; home working;

\section{Introduction}

The coronavirus pandemic (COVID-19) which started in 2019 quickly became a global crisis event, resulting in the mass quarantine of 100s of millions of citizens across numerous countries around the world. At the time of writing, the World Health Organisation (WHO) Coronavirus Disease (COVID-19) Dashboard reported over 7.5 million confirmed cases and in excess of 430,241 deaths\cite{whodash2020} globally. 
As COVID-19 spread across the globe, it also led to a secondary significant threat to a technology-driven society; i.e., a series of indiscriminate, and also a set of targeted, cyber-attacks and cyber-crime campaigns. Since the outbreak, there have been reports of scams impersonating public authorities (e.g., WHO) and organisations~ (e.g., supermarkets, airlines)\cite{malwarebytes2020cybereb,times2020fraud}, targeting support platforms\cite{krebsonsec2020cmap,TableGuardianNHSscam2020}, conducting Personal Protection Equipment (PPE) fraud\cite{europol-pandemic-profiteering1} and offering COVID-19 cures\cite{Norton2020EmailScam,Guardian2020EmailScam}. These scams target members of the public generally, as well as the millions of individuals working from home. Working at home \textit{en-masse} has realised a level of cyber security concerns and challenges never faced before by industry and citizenry. cyber-criminals have used this opportunity to expand upon their attacks, using traditional trickery (e.g., \cite{nursecybercrime2019}) which also prays on the heightened stress, anxiety and worry facing individuals. In addition, the experiences of working at home revealed the general level of unpreparedness by software vendors, particularly as far as the security of their products was concerned.

Cyber-attacks have also targeted critical infrastructure such as healthcare services\cite{wired2020hackp}. In response to this, on April 8th 2020, the United Kingdom's National Cyber Security Centre (NCSC) and the United States Department of Homeland Security (DHS) Cybersecurity and Infrastructure Security Agency (CISA) published a joint advisory on how cyber-criminal and advanced persistent threat (APT) groups were exploiting the current COVID-19 pandemic\cite{ncscdhsadv2020}. This advisory discussed issues such as phishing, malware and communications platform (e.g., Zoom, Microsoft Teams) compromise. What is arguably lacking here and in research, however, is a broader assessment of the wide range of attacks related to the pandemic. The current state of the art is extremely dispersed, with attacks being reported from governments, the media, security organisations and incident teams. It is therefore extremely challenging for organisations to develop appropriate protection and response measures given the dynamic environment. 

In this paper we aim to support ongoing research by proposing a novel timeline of attacks related to the COVID-19 pandemic. This timeline and the subsequent analysis can assist in understanding those attacks and how they are crafted, and as a result, to better prepare to confront them if ever seen again. Our timeline maps key cyber-attacks across the world against the spread of the virus, and also measures such as when lockdowns were put in place. The timeline reveals a pattern which highlights cyber-attacks and campaigns which typically follow events such as announcements of policy. This allows us to track how quickly cyber-attacks and crimes were witnessed as compared to when the first pandemic cases were reported in the area; or, indeed, if attacks preempted any of these events. We expand the timeline to focus on how specific attacks unfolded, how they were crafted and their impact on the UK. To complement these analyses, we reflect more broadly on the range of attacks reported, how they have impacted the workforce and how the workforce may still be at risk. In many ways this timeline analysis also forms a key contribution of our work both in terms of the chronological sequencing of attacks and the representation of campaigns using an accepted attack taxonomy. This therefore provides a platform which aligns with current literature and also provides the foundation which other research can easily build on. 

This paper is structured as follows. Section~\ref{sec:litrev} reflects on relevant cyber-attack and cyber-crime literature, and considers how opportunistic attacks have emerged in the past due to real-life crises/incidents. We then present our COVID-19-related cyber-attack timeline in Section~\ref{sec:timeline} as well as a dedicated focus on the United Kingdom as a case study of key-cyber-criminal activity. This is followed by a broader reflection on the attacks (those within and outside of the timeline). In Section~\ref{ImpactOnWorkforce} we discuss the impact of attacks on those working from home and wider technology risk. Section~\ref{sec:conclusion} concludes the paper and outlines directions for future work.

\section{Literature review}\label{sec:litrev}

With the broad adoption of digital technologies many facets of society have moved online, from shopping and social interactions to business, industry, and unfortunately, also crime. The latest reports establish that cyber-crime is growing in frequency and severity\cite{hiscox-cyberreadiness2019}, with a prediction to reach \$6 trillion by 2021 (up from \$3 trillion in 2015)\cite{ventures-annual2019} and even take on traditional crime in number and cost\cite{cbs-neth2020, anderson2019measuring}. Due to its lucrative nature\cite{mcguire2018into} and low risk level (as cyber-criminals can launch attacks from anywhere across the globe), it is clear that cyber-crime is here to stay.

cyber-crime, as traditional crime, is often described by the crime triangle\cite{CrossMichael2008Sotc}, which specifies that for a cyber-crime to occur, three factors have to exist: a victim, a motive and an opportunity. The victim is the target of the attack, the motive is the aspect driving the criminal to commit the attack, and the opportunity is a chance for the crime to be committed (e.g., it can be an innate vulnerability in the system or an unprotected device). Other models in criminology, such as Routine Activity Theory (RAT)\cite{yar2005novelty} and the fraud triangle\cite{cressey1953other} use similar factors to describe crimes, with some replacing the victim by the means of the attacker, which it can be considered otherwise as part of the opportunity. 

While attacks today have become more sophisticated and targeted to specific victims depending on attacker's motivation, for example for financial gain, espionage, coercion or revenge; opportunistic untargeted attacks are also very prevalent. We define ``opportunistic attacks'' as attacks that select the victims based on their susceptibility to be attacked\cite{dhanjani2009hacking}. Opportunistic attackers pick-up victims that have specific vulnerabilities or use hooks, usually in the form of social engineering, to create those vulnerabilities. Thus, we define as \textit{hook} any mechanism used to mislead a victim into falling prey of an attack.

These hooks take advantage of distraction, time constraints, panic and other human factors to make them work\cite{stajano2011understanding,nursecybercrime2019}. When victims are distracted by what grabs their interest/attention or when they are panicked, they are more susceptible to be deceived. Similarly, time constraints put victims under more pressure which can lead to mistakes and an increased likelihood to fall victim to scams and attacks. Other examples include work pressure, personal change of situation, medical issues, or events that cause deep and traumatic impact in the whole society in general such as fatalities and catastrophes.

Opportunistic attackers always seek to maximise their gain, and therefore, will wait for the best time to launch an attack where conditions fit those mentioned above. A natural disaster, ongoing crisis or significant public event are perfect cases of these conditions\cite{tysiac-2018}. In the past, several opportunistic attacks have been observed that took advantage of specific incidents; below, we provide few examples:

\begin{itemize}
\item Natural disasters: In 2005 Hurricane Katrina caused massive destruction in the city of New Orleans and surrounding areas in the USA\cite{fbi-katrina}. Not long after, thousands of fraudulent websites appeared appealing for humanitarian donations, and local citizens received scam emails soliciting personal information to receive possible payouts or government relief efforts. Similar scams and attacks have been witnessed in countless natural disasters since, such as the earthquakes in Japan and Ecuador in 2016\cite{ftcecja2016}, Hurricane Harvey in 2017\cite{cnetharvey2017}, or the bush fires in Australia in 2020\cite{abcaustr2020}.

\item Notable incidents or events: On 25th June 2009, the tragic death of Michael Jackson dominated news around the world. Only 8 hours after his demise, spam emails claiming knowing the details of the incident were circulating online\cite{sophos-jackson}. Waves of illegitimate emails echoing the fatality appeared soon after, containing links promising access to unpublished videos and pictures or Jackson's merchandise, that in reality were linked to malicious websites, or emails with malware infected attachments\cite{crn-jackson}. Noteworthy public events also attract a range of cyber-crime activities. During the FIFA World Cup in 2018 for instance, there were various attempts to lure individuals with free tickets and giveaways\cite{eset2018fifa}. These were, in fact, scams leading to fraud. 

\item Security incidents: In 2012, 164 million of email addresses and passwords were exposed in a LinkedIn data breach\cite{linkedin-breach}. This data was not disclosed until 4 years later, 2016, when it appeared for sale in the dark market. Soon after that, opportunistic attackers began to launch a series of attacks. Many users experienced scams, such as blackmail and phishing, and some compromised accounts that had not changed their passwords since the breach, were used to send phishing links via private message and InMail\cite{linkedin-malwarebytes}. 

\end{itemize}

Considering the variety of scams and cyber-attacks occurring around the events above, it is unsurprising that similar attacks have emerged during the ongoing COVID-19 pandemic. The outbreak has caused mass disruption worldwide, with people having to adapt their daily routines to a new reality: working from home, lack of social interactions and physical activity, and fear of not being prepared\cite{anxiety-who, anxiety-nhs}. These situations can overwhelm many, and cause stress and anxiety that can increase the chances to be victim of an attack. Also, the sudden change of working contexts, has meant that companies have had to improvise new working structures, potentially leaving corporate assets less protected than before for the sake of interoperability. 

Since the COVID-19 started, the numbers of scams and malware attacks have significantly risen\cite{sophos-covid-myriad}, with phishing being reported to have increased by 600\% in March 2020\cite{barracuda-threat-spotlight}. During April 2020, Google reportedly blocked 18 million malware and phishing emails related to the virus daily\cite{google-protecting-businesses-covid}. To increase likelihood of success, these attacks target sale of goods in high demand (e.g., Personal Protection Equipment (PPE) and coronavirus testing kits and drugs), potentially highly profitable investments in stocks related to COVID-19, and impersonations of representatives of public authorities like WHO and aid scams\cite{europol-pandemic-profiteering1, bloomberg-covid-aid-scams}. 

Brute force attacks on the Microsoft Remote Desktop Protocol (RDP) systems have increased as well\cite{kaspersky-remote-spring}, signaling attacks also on technology, not only on human aspects. It is clear then that attackers are trying the make the most of the disruption caused by pandemic, particularly given it continues to persist. As a consequence, several guidelines and recommendations have also been published to protect against attacks\cite{guidance-gov-uk, guidance-nist, guidance-ftc}. These guidelines are imperative for mitigating the increasing threat, but to strengthen their basis, there first needs to be a core understanding of the cyber-attacks being launched. This paper seeks to address this gap in research and practice by defining a timeline of cyber-attacks and consideration of how they impact citizens and the workforce.

\section{Timeline of COVID-19 related cyber-attacks \label{sec:timeline}}

The cyber-crime incidents erupting from the COVID-19 pandemic pose serious threats to the safety and global economy of the world-wide population, hence understanding their mechanisms, as well as the propagation and reach of these threats is essential. Numerous solutions have been proposed in the literature to analyse how such events unfold ranging from formal definitions to systemic approaches reviewing the nature of threats\cite{tsakalidis2017systematic, kotenko2013cyber, hindy2018taxonomy}. While these approaches enable the categorisation of the attack, they often lack the ability to map larger, distributed events such as the ones presented in this manuscript, where numerous events stem from the pandemic are, however, unrelated. To this end, we opted for temporal visualisation, enabling us to map events without compromising the narrative\cite{kolomiyets2012extracting}. Furthermore, this type of visualisation is used across the cyber-security domain to represent consequent cyber-attacks\cite{van2016classification,horton2018sony, falliere2011w32}.

\subsection{Approach to timeline creation}

In this section, we outline the methodology used to create the timeline.  We explain the search terms used to gather relevant COVID-19 cyber-attack data, the data sources (search engines) utilised, the sources of information we chose to focus on, and types of attack.  We also acknowledge the potential limitations of the work. 

\subsubsection{Nomenclature}
We explore a range of cyber-attacks which have occurred during the COVID-19 pandemic.  The novel coronavirus has been referred to by several different terms in the English-speaking world, including Coronavirus, Covid19, COVID-19, 2019-nCoV, and SARS-CoV-2.  We use the term COVID-19 to refer to the virus, which falls in line with terminology used by the World Health Organization \cite{whovirusname2020}.

\begin{figure*}[!ht]
	\centering
	\includegraphics[width=.8\linewidth]{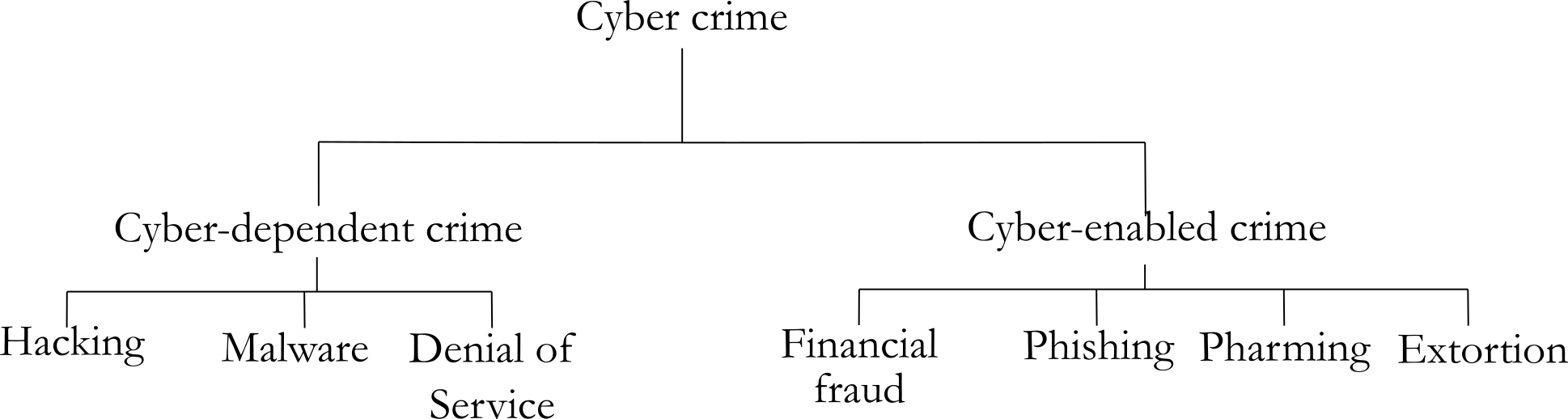}
	\caption{Cyber-dependent and cyber-enabled crimes\cite{CPS2019}}	\label{fig:cyber-crime}
\end{figure*}

\subsubsection{Construction of the timeline}
To aid in the construction of the timeline, we initially conducted a number of searches to identify cyber-attacks associated with the pandemic. These cyber-attacks were categorised by attack type, delivery method, and were ordered by date.  The information gathered has been collated and is presented in Figure \ref{fig:timeline} which serves as a baseline for the construction of Table~\ref{tableOfAttacks}.

Information presented in the timeline includes the date China alerted the WHO about the virus, the date the pandemic was officially declared, and cyber-attacks which specifically relate to hospitals or medicine. Additionally, key countries involved in the pandemic were identified, and for those, we present the first identified case, the date lockdown was implemented, and the first cyber-attack they suffered. The table seeks to examine a sub-set of the information from the timeline.

Furthermore, we have chosen to include a number of sources offering reports of attacks. The sources are a mixture of reputable news outlets (such as Reuters, and the BBC), blog articles, security company reports, and social media posts. Though blog articles and social media posts are not considered to be an academic source, in the context of this research where we are examining an emerging threat, they offer important insights into trends of cyber-attacks. It is also important to note that cyber-attacks may first be presented in these domains, before being highlighted by mainstream media outlets. With regards to the inclusion of news reports in the table of attacks and subsequent timeline, it should be acknowledged that these attacks are being presented through a journalistic lens, and as such may be written in an attempt to grab headlines. Nevertheless, these reported cyber-attacks still pose a tangible threat to the general public during the COVID-19 pandemic. The timeline seeks to provide an overview of attacks which have occurred.

The state-of-the-art review of reports was performed from mid-March to mid-May 2020.  The timeline limits cyber-attacks to those experienced by 31st March. This is because we reached what we believed to be a saturation point comprising a sufficient number of cyber-attacks to be representative.  Following the conclusion of the search, the earliest reported attack was on 6th January 2020 \cite{TableFireAPT322020}, whilst the most recently listed attack in the timeline was 31st March 2020 \cite{TableThreatSkype2020}. The most recently listed attack in the table was 13th May 2020 \cite{Ng2020}. The table progresses the time period a bit further as it intends to provide more detail in regards to cyber-attacks experienced during this time. Sources were gathered from a number of locations.  The criteria used to locate reports have been defined below and are presented in a similar way to existing reviews in cyber security literature\cite{shepherd2018, chockalingam2017bayesian}. The structure of the timeline is described in further detail in Section \ref{TheTimeline}.

\textbf{Search engines:} Several search engines were used in the creation of the table and timeline.  These were- Google\footnote{www.google.com} (US-based and dominates the search engine market share), Baidu\footnote{www.baidu.com} (Chinese-based search provider), Qwant\footnote{www.qwant.com} (French-based search engine with a focus on privacy), and DuckDuckGo\footnote{www.duckduckgo.com} (US-based search engine with a focus on privacy).

\begin{CJK*}{UTF8}{gbsn}
\textbf{Keywords used:} A variety of keywords were used when collating reports of cyber-attacks.  Non-English terms were translated using the Google Translate service \cite{googletranslate2020} and additional independent sources were used as a means of validating the translation.  When focussing on the virus itself, the following key words were used: sars-cov-2, Covid, Covid19, Coronavirus, 冠状病毒 (Chinese translation for Coronavirus, confirmed by the World Health Organization\cite{whochinese2020}), コロナウイルス (Japanese translation for Coronavirus, confirmed by the Japanese Ministry of Health, Labour and Welfare\cite{japanministry2020}). 

When searching for cyber-attacks, the following key phrases were used: 网络攻击  (Chinese translation means Network Attack\cite{wangxian2009research} or Cyber Attack\cite{WhoFive2020}), サイバー攻撃 (Japanese translation for Cyber Attack or Hacking Attack \cite{jisho2020}), Attaque Informatique (French translation for Computer Attack \cite{Parisien2020}), Attacco Informatico (Italian translation for Cyber Attack \cite{Repubblica2020}).  
\end{CJK*}

\textbf{Time range:} We attempted to find the earliest reported cyber-attack which was associated with the COVID-19 pandemic. To allow for development of the timeline, and analysis of findings, mid-May 2020 was defined as a cut-off point, with the most recent news article being dated 13th May 2020\cite{TableGuardianNHSscam2020}.

\textbf{Exclusion criteria:} Although we have created a comprehensive table and timeline, a number of results were excluded from the research. These included results which a) were behind a paywall, b) required account creation before full article was displayed, c) were duplicates of existing news reports, and d) could not be translated.

\subsubsection{Types of cyber-attacks \label{TypesOfCyberAttacks}}
To guide our analysis and the creation of a timeline of COVID-19-related cyber-attacks, we decided to define attacks based on their types. This allowed us to examine the prominence in certain types of attacks. Although there exist numerous taxonomies relating to attacks and cyber-crimes (e.g., \cite{Ciardhuain2004,nursecybercrime2019,Cebula2010,Nicholson2012}), there exists no universally accepted model\cite{9108270}. In this work therefore, we relied on the UK's Crown Prosecution Service (CPS) categorisation of cyber-crime. This definition includes cyber security by default and has inspired many international definitions of cyber-crime.

The CPS guidelines categorise cyber-crime into two broad categories: \textit{cyber-dependent} and \textit{cyber-enabled} crimes\cite{CPS2019}. A cyber-dependent crime is an offence, \textit{``that can only be committed using a computer, computer networks or other form of information communications technology (ICT)''}\cite{Mcguire2013CDC}. Cyber-enabled crimes are, \textit{``traditional crimes, which can be increased in their scale or reach by use of computers, computer networks or other forms of information communications technology (ICT)''}\cite{Mcguire2013CEC}. These categories as well as examples of their subcategories can be seen in Figure~\ref{fig:cyber-crime}. Some of the elements described by CPS are often interlinked in a cyber-attack. For instance, a phishing email or text message (e.g., SMS or WhatsApp) might be used to lure a victim to a fraudulent website. The website then may gather personal data which is used to commit financial fraud, or it may install malware (more specifically, ransomware) which is then used to commit extortion. This notion of cyber-attack sequences is explained in further detail in Section \ref{TheTimeline}.

Similarly Denial of Service (DoS) attacks are increasingly used by cyber-criminals to distract (or, act as `smokescreens' for) businesses during hacking attempts\cite{kaspersky2016ddossc,bellekens2019cyber}. In what follows, we consider the types of these attacks and reflect on how they have been launched, including any human factors or technical aspects (e.g., vulnerabilities) they attempt to exploit. 

Phishing, or Social Engineering more broadly, includes attempts by illegitimate parties to convince individuals to perform an action (e.g., share information or visit a website) under the pretence that they are engaging with a legitimate party. Quite often email messages are used, occasionally SMS or WhatsApp messages are used (referred to as smishing). Pharming is similar to phishing but instead of deceiving users into visiting malicious sites, attackers rely on compromising systems (e.g., the user's device or DNS servers) to redirect individuals to illegitimate sites. This type of attack is less common in general, as it requires more access or technical capabilities. Financial fraud generally involves deceiving individuals or organisations using technology for some financial gain to the attacker or criminal. Extortion refers to actions that force, threaten or coerce individuals to perform some actions, most commonly, releasing finances. 

Hacking, Malware and Denial of Service (DoS) attacks are forms of crime that are often favoured by more technical attackers. Hacking involves compromising the confidentiality or integrity of a system, and requires a reasonable about of skill; its techniques can involve exploiting system vulnerabilities to break into systems. Malware refers to malicious software and can be used for disrupting services, extracting data and a range of other attacks. Ransomware is one of the most common type of malware today\cite{CSO2020jf,malwarebytes2020smr}, and combines malware with extortion attempts. DoS attacks target system availability and work by flooding key services with illegitimate requests. The goal here is to consume the bandwidth used for legitimate server requests, and eventually force the server offline. 

These types of attack provide the foundation for our analysis in the timeline and how we approach our discussion in later part of this research.

\subsubsection{Limitations of the table}
Within Table~\ref{tableOfAttacks}, two columns referring to dates are provided.  The first column ``Article Date'' refers to the date the reference was initially published. We acknowledge that in some cases, the web pages linked to the references continued to be updated with information following its inclusion with the paper. The table has been ordered by ``Article Date" to provide a consistent chronological representation of events.  

We have also provided a second column,``Attack Date''. When examining each reference, if a specific date was provided as to when the attack was executed, it was  included. The rational behind including the attack date and report date is that an attack may not surface until several days after it has been carried out.

\subsubsection{Limitations of the timeline}
Two types of cyber-attack reports are  considered within this manuscript, those which describe cyber-attacks without providing the date of the attack and those which describe cyber-attacks and include the date of the perpetration. When the date of the attack is not included, the date provided in the timeline refers to the date of the publication. The rationale behind the inclusion of both types of reports is based on providing a chronological representation of events. Furthermore, while the table provides an extensive overview of the threat landscape, it is by no means an exhaustive list of all the attacks carried out in relation to the pandemic, as gathering such information would not be possible in this context due to the lack and quality of reporting, the number of targeted incidents, the number of incidents targeted at the general public, the global coverage of the pandemic and the number of malicious actors carrying out these attacks. 

However, despite these limitations we have explored all resources available to depict the threat landscape as accurately as possible. 

\subsection{The timeline}\label{TheTimeline}

\begin{sidewaysfigure*}
    \centering
    \includegraphics[width=1\linewidth]{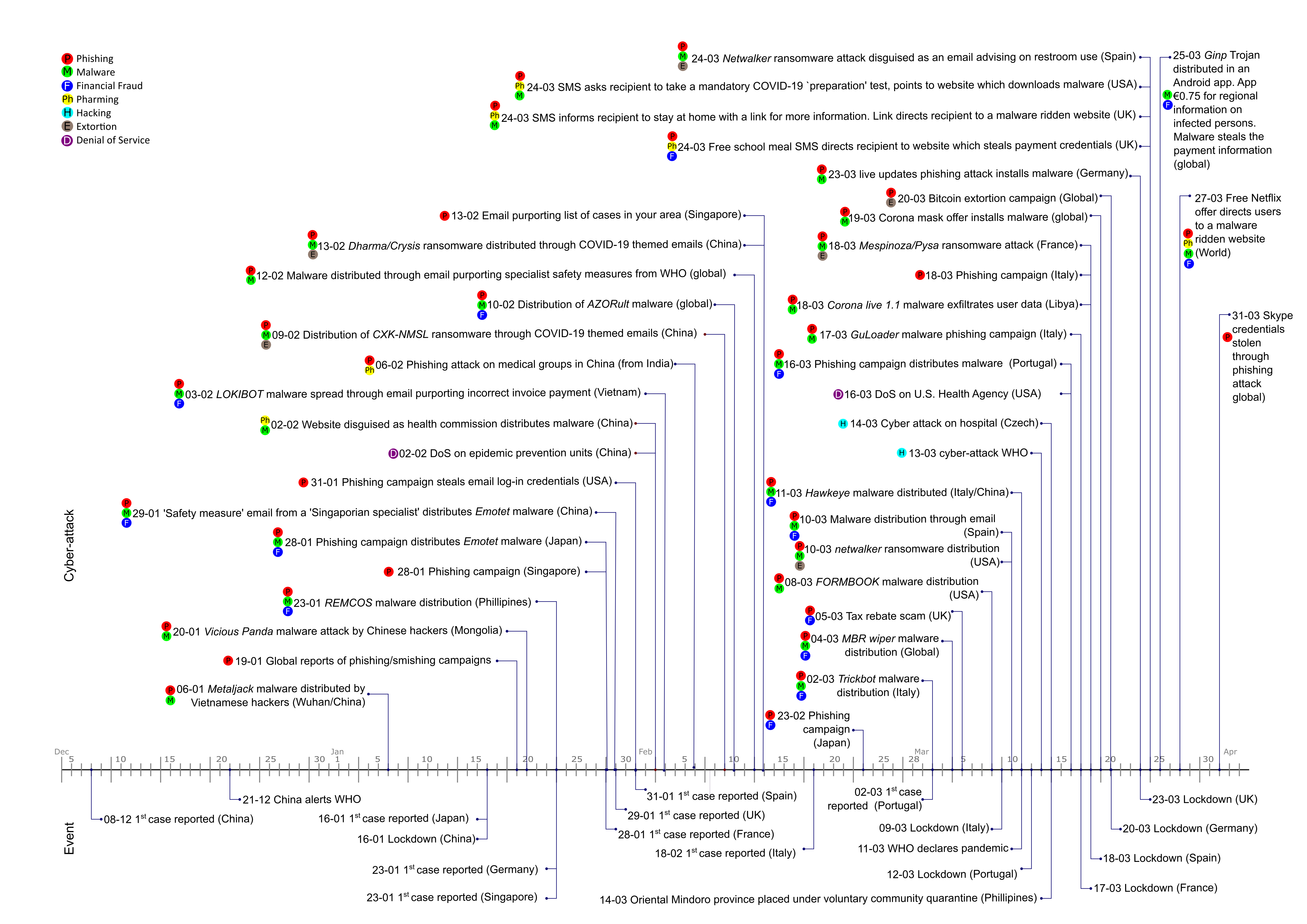}
    \caption{Timeline of Key Events Related to Cyber-Attacks and the COVID-19 Pandemic}
    \label{fig:timeline}
\end{sidewaysfigure*}

In this section, we examine the cyber-attacks in further detail. Figure~\ref{fig:timeline} provides a detailed temporal representation of the chain of key cyber-attacks induced by the COVID-19 pandemic. The timeline includes the first reported cases in China, Japan, Germany, Singapore, Spain, UK, France, Italy, and Portugal and then the subsequent lockdown announcements. The timeline presents 43 cyber-attacks categorised using the CPS taxonomy described in Section \ref{TypesOfCyberAttacks} and abbreviated as: \textit{P:phishing} (or \textit{smishing}), \textit{M:malware. Ph:pharming, E:extortion, H:hacking, D:denial of service} and \textit{F:financial fraud.} The events related to the crisis were validated against WHO timeline of events to ensure an accurate temporal reproduction.

Table \ref{tableOfAttacks} describes a number of cyber-attacks in further detail. Within the table, cyber-attacks have been organised by attack date.  If the attack date was not available within the reference, then the article date has been used.  The target-country of each cyber-attack has been listed, alongside a brief description of the methods involved.  Finally, the attack type has also been classified in accordance with the CPS taxonomy described earlier where it has been mentioned within the reference. 

Both the figure and the table present specific cyber-attacks and incidents and exclude: general advisories (e.g. from governmental departments), general discussions and summaries of attacks, and detailed explanations of techniques and approaches utilised by the attackers.

\begin{table*}
\fontsize{7}{9}\selectfont
\caption{Descriptions of COVID-19 related cyber-attacks \label{tableOfAttacks}}
\begin{tabular}{|p{0.4cm}|p{0.5cm}|p{1.1cm}|p{1cm}|p{9cm}|p{0.8cm}|p{0.8cm}|}

\hline 

\rowcolor{gray!50} \textbf{ID} &\textbf{Ref.} &  \textbf{Country} &  \textbf{Attack Type} & \textbf{Description} & \textbf{Article Date} & \textbf{Attack Date} \\ \hline

\rownumber & \cite{TableFireAPT322020} & China & P.M & Vietnam accused of launching a \textit{METALJACK} phishing campaign against the Wuhan district offices & 22/04 & 06/01 \\ \hline

\rownumber & \cite{TableAON2020} & Global & P.M & International reports that both phishing and smishing campaigns are taking place& 19/01 & - \\ \hline

\rownumber & \cite{TableForbes2020} & China, Mongolia & P.M & Chinese hackers accused of distributing the \textit{Vicious Panda} malware to Mongolia through emails purporting to come from the Mongolian ministry of affairs & 12/03 & 20/01 \\ \hline

\rownumber & \cite{TableFSec2020} & Phillipines & P.M.F & \textit{REMCOS} malware distributed to Phillipino citizens & 13/03 & 23/01 \\ \hline

\rownumber & \cite{TableKaspersky2020} & Singapore & P & Phishing campaign steals email log-in credentials & 28/01 & - \\ \hline

\rownumber & \cite{SentinelOne2020} & Japan & P.M.F & Safety measures phishing campaign distributes \textit{Emotet} malware & 28/01 & 28/01 \\ \hline

\rownumber & \cite{Tablesmzdm2020} & China & P.M.F & 'Safety measure' email from a 'Singaporian specialist' distributes \textit{Emotet} malware & 06/02 & 29/01 \\ \hline

\rownumber & \cite{TableKaspersky2020} & USA & P & Email purporting list of COVID-19 cases in victim's city takes user to website which steals credentials & 11/02 & 31/01\\ \hline

\rownumber & \cite{TableCSDN2020} & China & H & DoS on epidemic prevention units & 09/02 & 02/02 \\ \hline

\rownumber & \cite{TableCSDN2020} & China & P & Phishing campaign steals email log-in credentials & 09/02 & 02/02 \\ \hline

\rownumber & \cite{TechRep2020} & World & P.M.F & First cases of \textit{AZORult} a data theft malware & 10/02 & - \\ \hline

\rownumber & \cite{Tablecqgbxa2020} & China & P.M & Email purporting specialist safety measures from WHO prompts malware download & 12/02 & - \\ \hline

\rownumber & \cite{TableFSec2020} & Vietnam & P.M & \textit{LOKIBOT} malware spread through email purporting incorrect invoice payment & 13/03 & 03/02 \\ \hline

\rownumber & \cite{Patranobis2020} & China & P.Ph & Phishing attack on medical groups in China (from India) & 06/02 & 06/02 \\ \hline

\rownumber & \cite{Tablefreebuf2020} & China & P.M.E & Distribution of \textit{CXK-NMSL} ransomware through COVID-19 themed emails & 18/02 & 09/02 \\ \hline

\rownumber & \cite{Tablefreebuf2020} & China & P.M.E & Distribution of \textit{Dharma/Crysis} ransomware through COVID-19 themed emails & 18/02 & 13/02 \\ \hline

\rownumber & \cite{TableFSec2020} & Italy & P.M & \textit{Trickbot} malware distributed through email  & 13/03 & 02/03 \\ \hline

\rownumber & \cite{Stonefly2020} & Global & P.M.F & MBR wiper malware disguised as contact tracing information & 04/03 & - \\ \hline

\rownumber & \cite{TableFSec2020} & USA & P.M & \textit{FORMBOOK} malware distributed through email purporting parcel shipment advice & 13/03 & 08/03 \\ \hline

\rownumber & \cite{TableRegister2020} & USA & M & Health systems in Champaign Urbana Public Health District (Illinois) affected by the \textit{netwalker} ransomware & 12/03 & 10/03 \\ \hline

\rownumber & \cite{TableFSec2020} & Spain & P.M & Email purports COVID-19 remedy as mooted by Israeli scientists days in advance & 13/03 & 10/03 \\ \hline

\rownumber & \cite{TableSCMedia2020} & Czech & H & Cyber-attack on Czech hospital  & 14/03 & 14/03 \\ \hline

\rownumber & \cite{TableBloomberg2020} & USA & H & Denial of Service on U.S. Health Agency & 16/03 & - \\ \hline

\rownumber & \cite{TableLookout2020} & Libya & P.M & Corona live 1.1 is the \textit{SpyMax} malware which in this case is a trojanised app which exfiltrates user data & 18/03 & - \\ \hline

\rownumber & \cite{TableZscaler2020} & World & P.M & Corona mask offer installs what appears to be a harmless malware which distributes an SMS to all contacts. Presumably an update to the app will mobilise the malware & 19/03 & - \\ \hline

\rownumber & \cite{TableESET2020} & Global & P.E & Extortion campaign threatens to infect the recipient with COVID-19 unless a \$4,000 bitcoin payment is made & 17/04 & 20/03 \\ \hline

\rownumber & \cite{TableMurica2020} & Spain & P.M & \textit{Netwalker} ransomware attack disguised as an email advising on restroom use & 24/03 & - \\ \hline

\rownumber & \cite{TableTwitterBen2020} & USA & P.M & SMS asks recipient to take a mandatory COVID-19 `preparation' test, points to website which downloads malware & 24/03 & 24/03 \\ \hline

\rownumber & \cite{TableTwitterGlos2020} & UK & P.M & SMS informs recipient to stay at home with a link for more information. Link directs recipient to a malware ridden website & 24/03 & - \\ \hline

\rownumber & \cite{TableBirminghamMail2020} & UK & P.Ph.F & Free school meal SMS directs recipient to website which steals payment credentials & 25/03 & 24-03 \\ \hline

\rownumber & \cite{TableInfosecurity2020} & World & M.F & \textit{Ginp} Trojan distributed in an Android app. App charges \euro{0.75} for information on infected persons in the recipients region. In actual fact, it steals the payment information & 25/03 & - \\ \hline

\rownumber & \cite{TableThreatSkype2020} & Global & P & Skype credentials stolen through a crafted phishing campaign & 23/04 & 31/03 \\ \hline
 
\rownumber & \cite{TableMetaCompliance2020} & World & P.Ph.M.F & Free Netflix offer directs users to a malware ridden website & 27/03 & - \\ \hline

\rownumber & \cite{TableDailyMail2020} & UK & M & Fake NHS website gathers user credentials & 28/04 & -  \\ \hline

\rownumber & \cite{TableDataReview2020} & UK & P.M & Email purports to offer job retention payment as per the UK governmental announcement & 30/04 & 19/04  \\ \hline

\rownumber & \cite{TableBleepScreen2020} & Global & M & \textit{Coronalocker} locks a computer and appears to cause rather more annoyance than any real damage & 21/04 & - \\ \hline

\rownumber & \cite{TableDarkReading2020} & Global & P.M & Docusign recipients directed to fake website offering COVID-19 information & 08/05 & - \\ \hline
 
\rownumber & \cite{TableGuardianNHSscam2020} & UK & P.M & Recipients are directed to a fake track and trace website which collects user credentials & 13/05 & - \\ \hline

\multicolumn{7}{p{14cm}}{key: P:Phishing (or smishing); Ph:Pharming; E:Extortion; M:Malware; F:Financial fraud; H:Hacking}
\end{tabular}
\end{table*}

\normalsize

\subsection{COVID-19 cyber-attacks in the United Kingdom  \label{AttackCaseStudy}} 

The extent of the cyber-security related problems faced in the UK was quite exceptional, and in this section we use the UK as a case study to analyse COVID-19 related cyber-crime. The discussion herein demonstrates that as expected and outlined above, there was a loose correlation between policy/news announcements and associated cyber-crime campaigns. The analysis presented herein focuses only on cyber-crime events specific to the UK. So for example, although many of the incidents identified in the previous section and particularly in \cite{Mimecast2020} are global cyber-attacks, the discussion herein ignores these. Consequently, numerous announcements purportedly coming from reputed organisations such as WHO and a plethora of malware which reached UK citizens is ignored as these were not UK specific issues.

Indications of the extent of the UK cyber-crime incident problem experienced during the pandemic are provided by the reported level of suspect emails and fraud reported. By early May (07-05-20), more than 160,000 `suspect' emails had been reported to the National Cyber Security Centre \cite{NCSC2020a} and by the end of May (29-05-20), £4.6m had been lost to COVID-19 related scams with around 11,206 victims of phishing and / or smishing campaigns \cite{SkyNews2020}. In response, the National Cyber Security Centre (NCSC) took down 471 fake online shops \cite{TableJoeBBC2020} and HMRC (Her Majesty's Revenue and Customs) took down 292 fake websites \cite{Infosecurity2020}.

\begin{sidewaysfigure*}
	\centering
	\includegraphics[width=1\linewidth]{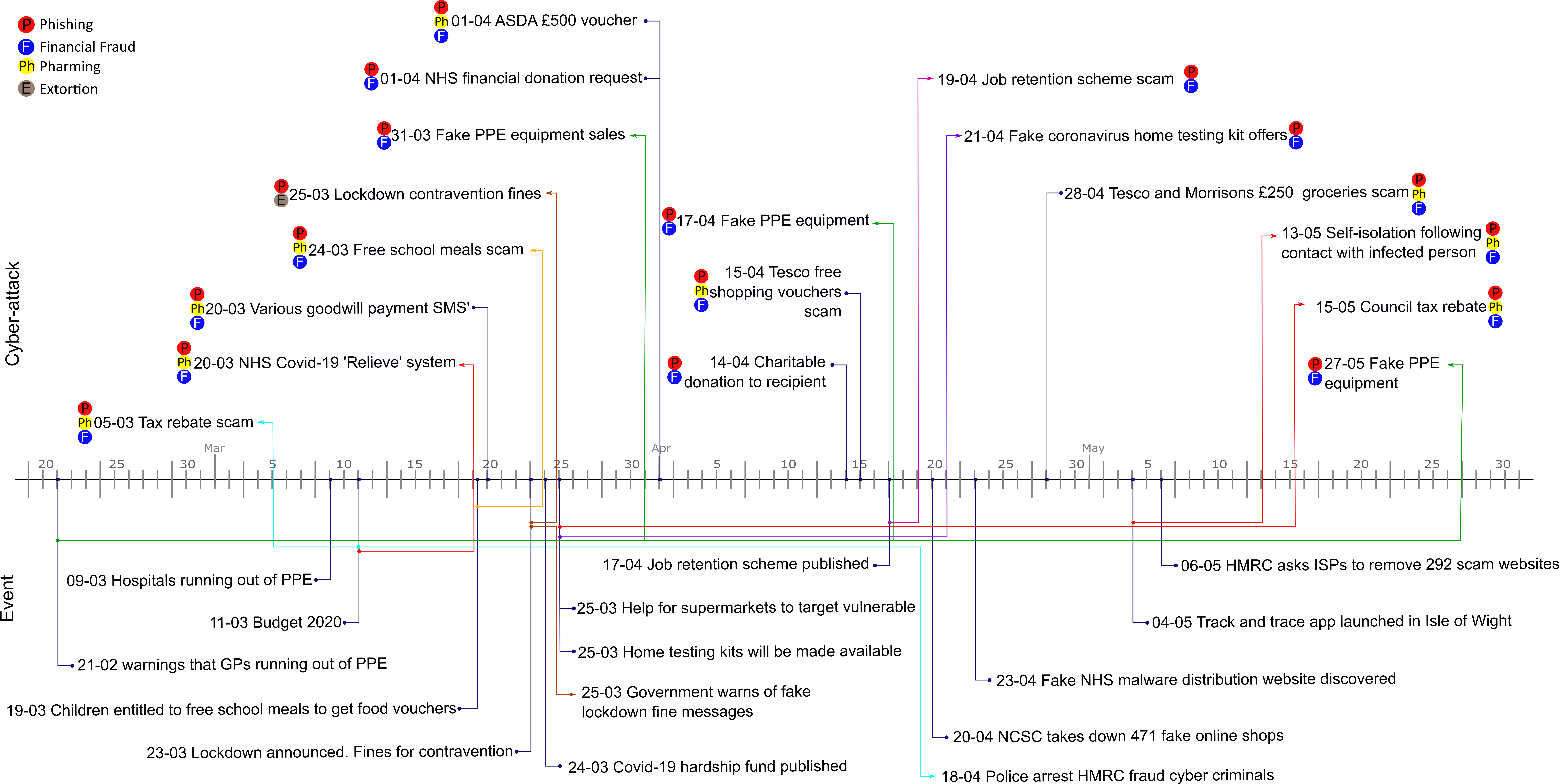}
	\caption{UK Timeline}
	\label{UKTimeline}
\end{sidewaysfigure*}

The timeline in Figure \ref{UKTimeline} shows a series of UK specific events and cyber-crime incidents. The timeline indicates a direct and inverse correlation between announcements and incidents.

\textit{Direct correlations} are instances where perpetrators appear to follow announcements or events, they may have drawn on these events and carefully configured cyber-attacks around policy context. These are shown in the figure with a solid coloured connecting arrow.

\textit{Inverse correlations} are instances where an incident has no clear correlation with an event or announcement. Although inverse correlations do not appear to have a direct correlation, these may exist because a number of events were being actively highlighted in the media. For example, the issue of personal protective equipment (PPE) was in active discussion well before the UK government gave this priority consideration. Similarly, the likelihood of a tax rebate scheme was in active consideration in early March before the budget announcement on 11-03-20. The first tax rebate phishing campaigns were in active circulation before the budget announcement. In both cases, we should emphasise that these are loose correlations and more work needs to be done in terms of whether a predictive model can be built using this data and data around the world as examples.

On 11th March 2020, the UK government made a number of important budgetary announcements \cite{UKGov2020a} which included: a £5bn emergency response fund to support the NHS and other public services in England; an entitlement to statutory sick pay for individuals advised to self-isolate; a contributory Employment Support Allowance for self-employed workers; a £500m hardship fund for councils to help the most vulnerable in their areas; a COVID-19 Business Interruption Loan Scheme for small firms; and the abolishment of business rates for certain companies. 

Soon after, the government continued to make announcements to support the citizenry and economy. These announcements included: a scheme to support children entitled to receive free school meals (19-03-20); a hardship fund (24-03-20); help for supermarkets to target vulnerable people (25-03-20); the potential availability of home test kits (25-03-20); a job retention scheme (17-04-20); and the launch of the much awaited \textit{track and trace} app (04-05-20). 

\begin{table*}[!t]
	\centering 
	\footnotesize
	\caption{Selected correlations between events and cyber-criminal campaigns \label{Correlation}}
	\begin{tabular}{|p{1.3cm}|p{5.4cm}|p{2.3cm}|p{5.4cm}|}
		\hline \hline
		\textbf{Event date} & \textbf{Event} & \textbf{Incident date \& type} & \textbf{Incident} \\ \hline  \hline
		21-02-20, 09-03-20 & Doctors warn GPs are running out of PPE; Hospitals running out of PPE & 17-04-20 \texttt{p,ph,f} \newline 27-05-20 \texttt{p,ph,f} & Fake PPE offers through email. Link to URLs which capture credit card and other details \\ \hline
		11-03-20 & Government announces a range of financial assistance packages in the budget & 20-03-20 \texttt{p,ph,f} & Smishing campaign promising a COVID-19 financial relief payment. Respondents are directed to a fake \textit{gov.uk} website which requests credit/debit card details \\ \hline
		19-03-20 & Government announces a scheme which entitles children who qualify for a free school meal to a food voucher or alternatives if they are not able to continue attending school. & 24-03-20 \texttt{p,ph,f} & A smishing campaign which targeted parents with a promise of help with their free school meals in return for banking details. Banking details are defrauded \\ \hline
		23-03-20 & Lockdown announced. £60 contravention fine, later (10-05-20) increased to £100 & 27-03-20 \texttt{p,e} & Lockdown contravention SMS \\ \hline
        24-03-20 & COVID-19 hardship fund enables councils to reduce council tax bills by £150 for residents of working age and who have had their bill reduced by an award of council tax reduction & 15-05-20 \texttt{p,ph,f} & Council tax rebate scam\\ \hline
        25-03-20 & Government announce intention to make home testing kits available & 31-03-20 \texttt{p,f}, \newline17-04-20 \texttt{p,f}, \newline 27-05-20 \texttt{p,f} & Phishing campaigns in England and Scotland direct victims to fake websites which claim to sell  PPE equipment\\ \hline
		17-04-20 & Government announces job retention scheme & 19-04-20 \texttt{p,f} & Fake job retention scheme phishing campaign. \\ \hline	\hline 
	\end{tabular}
\end{table*}

Events such as these increase the likelihood of a positive response to a cyber-criminal campaign and perpetrators are very likely to hook onto events. Although there appears to be a link between some of the events and incidents, a number of scams cannot easily be traced to a single event or announcement. Examples of this include a \textit{goodwill} payment of £250 (21-03-20), an NHS financial donation request (02-04-20), vouchers for UK supermarkets (02-04-20, 15-04-20, 28-04-20), and a charitable donation to the recipient. None of these events have associated governmental announcements or even general public speculation.

\begin{figure}[!b]
	\centering
	\includegraphics[width=1\linewidth]{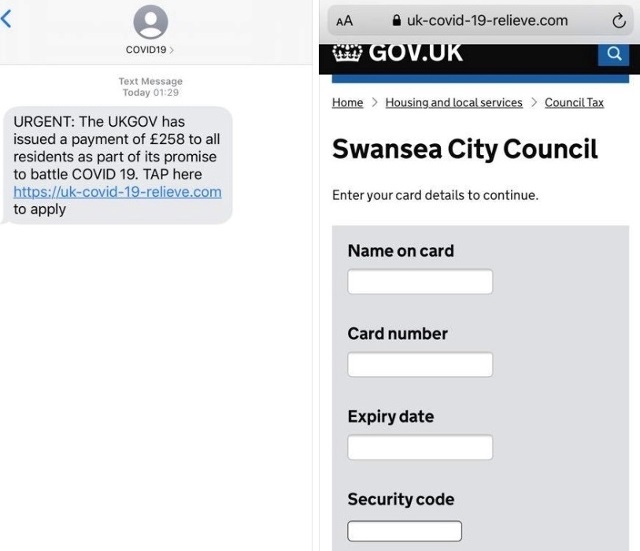}
	\caption{The COVID-19-relieve scam\cite{Swansea2020}}
	\label{CovidRelieveSMS}
\end{figure}

Examples supporting our notion of a correlation between events and cyber-security campaigns are provided in Table \ref{Correlation} and illustrated in Figure \ref{UKTimeline}. These examples indicate a loose correlation between events and cyber-criminal campaigns. Many of the cases outlined in Table \ref{Correlation} and Figure \ref{UKTimeline}, were very simple. Potential victims were provided URLs through email, SMS, or Whatsapp. An example of this is provided in Figure \ref{CovidRelieveSMS}. In this case, the URL pointed to a fake institutional website which requests credit/debit card details. Although there are elements of this process which are obviously suspicious to a more experienced computer user, for example, spelling errors (\textit{relieve} instead of \textit{relief} in the COVID-19 relief scam), suspect reply email addresses and clearly incorrect URLs, these are not immediately obvious to many users.

\subsection{Analysis of cyber-attacks and associated risks}\label{AnalysisOfTimelineEvents}

The timeline shown in Figure \ref{fig:timeline} and the UK case study above creates an ideal platform through which to analyse the cyber-attacks that have occurred in light of the pandemic.
From the point that the first case was announced in China (08-12-19), the first reported COVID-19 inspired cyber-attack took 30 days. The next reported cyber-attack was 14 days (19-01-20). From this point onwards, it is clear that the timeframe between events and cyber-attacks reduces dramatically. 

The 43 cyber-attacks presented in the timeline can be further categorised as follows: 

\begin{itemize}\setlength{\parskip}{0pt}\setlength{\itemsep}{0pt plus 2pt}
    \item 37 (86\%) involved phishing and / or smishing 
    \item 2 (5\%) involved hacking
    \item 2 (5\%) involved denial of service
    \item 28 (65\%) involved malware
    \item 15 (34\%) involved financial fraud
    \item 6 (13\%) involved pharming
    \item 6 (15\%) involved extortion
\end{itemize}

Whilst this analysis is useful, the sequence of events in the complete attack can also provide key attack insights. The timeline reveals these sequences and shows the complete campaign comprising of, for instance, the distribution of malware (\textit{m}) through phishing (\textit{p}) which steals payment credentials which are used for financial fraud (\textit{f}). We can describe this cyber-attack sequence as \texttt{p,m,f}. Analysing cyber-attacks in this way is important because this indicates multiple points in a cyber-attack where protections could be applied. 
The timeline reveals the following cyber-attack sequences:

\begin{itemize}\setlength{\parskip}{0pt}\setlength{\itemsep}{0pt plus 2pt}
    \item \texttt{p,m}: \textit{n=8, 19\%} 
    \item \texttt{p,m,f}: \textit{n=10, 23\%}
    \item \texttt{ph,m}: \textit{n=1, 2\%}
    \item \texttt{p,ph}: \textit{n=1, 2\%}
    \item \texttt{p.m.e}: \textit{n=5, 12\%}
    \item \texttt{p,ph,m}: \textit{n=2, 5\%}
    \item \texttt{p,ph,f}: \textit{n=1, 2\%}
    \item \texttt{p,e}: \textit{n=1, 2\%}
    \item \texttt{p,ph,m,f}: \textit{n=1, 2\%}
\end{itemize}

This analysis does not include the sequence of events that took place in the two hacking and two denial of service incidents. It should be noted that although financial fraud is the most likely goal in most of the cyber-attacks described in the timeline, financial fraud was only recorded in the timeline where reports have clearly indicated that this was the outcome of a cyber-attack. In reality, the \texttt{p,m,f} and \texttt{p,ph,f} cases are likely to be higher.

\begin{figure}[!b]
		\centering
		\includegraphics[width=1\linewidth]{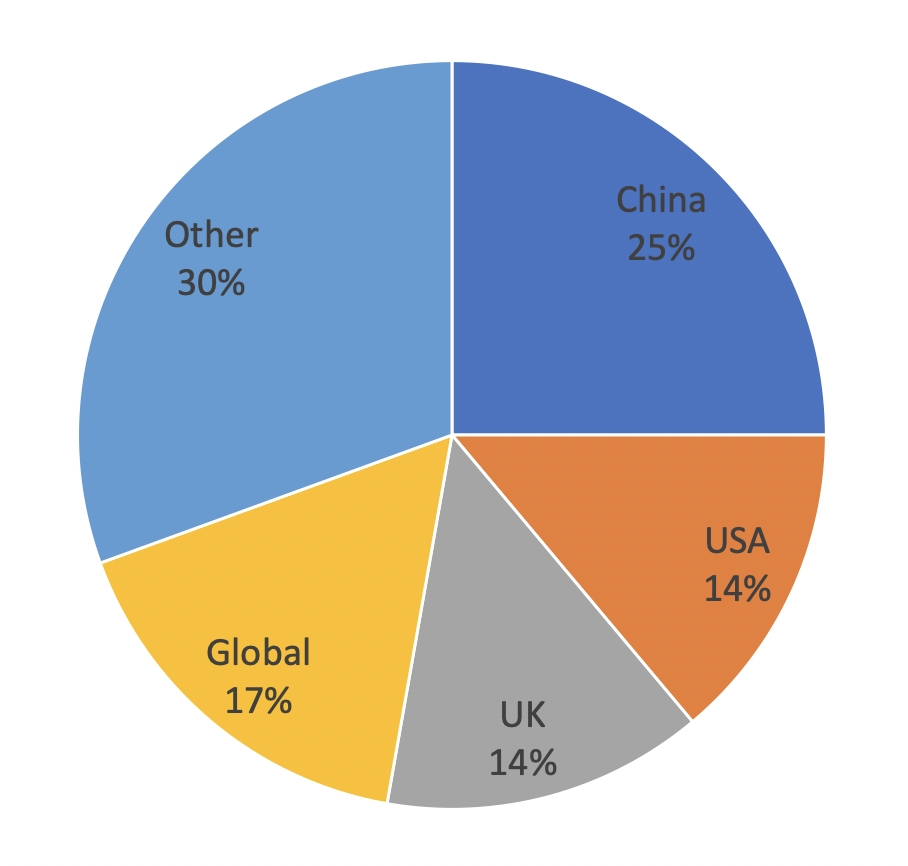}
        \caption{Cyber-attack distribution across countries examined}
        \label{fig:CountryLandscape}
\end{figure}

Figure~\ref{fig:CountryLandscape} provides a summary of the countries that were the target of early cyber-attacks during the pandemic, organised by attack date. As shown, China and the USA account for 39\% of the attacks reported. It is also clear from Table~\ref{tableOfAttacks} that both of these countries were primary target from the start of the pandemic. The attacks then spread to the United Kingdom and more other countries. By March 2020 however, a vast majority of the attacks are targeted at the whole world, with a reminder of attacks specifically focused at events in a single country, such as tax rebates due to COVID-19, or contact tracing phishing messages.

It is useful to consider this in the context of UK specific cyber-attacks. This examination reveals that phishing was a component of all (\textit{n=17}) the cyber-attacks analysed. 1 involved extortion as the final goal, the remaining 16 involved financial fraud. 9 cyber-attacks comprised the sequence: \texttt{p,ph,f}, 7 comprised the sequence \texttt{p,f}, the remaining 1 comprised of \texttt{p,e}.

It is notable that although an NHS malware distribution website was discovered and removed on 23-04, none of the cyber-attacks we analysed appeared to involve malware in the same way that the global analysis reveals. There may be a number of reasons for this. Launching a malware connected campaign requires more sophistication and time. There may be less opportunity to directly connect it to a specific event or announcement. The time delay between some of the announcements and the associated campaigns was remarkably short. For instance, the time delay between the lockdown announcement (23-03-20) and the `lockdown contravention fine' (25-03-20) was 2 days, and the time delay between the job retention scheme announcement (17-04-20) and the job retention scam (19-04-20) was also 2 days.

To reflect more generally on the cyber-attacks discovered, we can see that phishing (including smishing) were, by far, the most common based on our analysis. In total, it was involved in 86\% of the global attacks. This is however, unsurprising, as phishing attempts are low in cost and have reasonable success rates. In the case of COVID-19, these included attempts at impersonating government organisations, the WHO, the UK's National Health Service (NHS), airlines, supermarkets and communication technology providers. The specific context of the attacks can be slightly different however the underlying techniques, and the end goal is identical. 

For instance, in one email impersonating the WHO, attackers attach a zip file which they claim contains an e-book that provides, ``\textit{the complete research/origin of the corona-virus and the recommended guide to follow to protect yourselves and others}''\cite{bellekens2016pervasive}\cite{malwarebytes2020cybereb}. Moreover, they state: ``\textit{You are now receiving this email because your life count as everyone lives count}''. Here, attackers are using the branding of WHO, posing as helpful (the remainder of the email contains legitimate guidance), and appealing to people's emotions in crafting their attack email\cite{Iuga2016,nursecybercrime2019}. Similar techniques can be seen in a fake NHS website created by criminals detected online, which possesses identical branding but is riddled with malware\cite{dailymail2020cybnhs}, and a malicious website containing malware which also presents the legitimate Johns Hopkins University COVID-19 dashboard\cite{krebsonsec2020cmap}. It is notable that the fake WHO email contains spelling/grammatical errors. The discussion in Section \ref{AttackCaseStudy} provides further specific examples of this. 

To further increase the likely success of phishing attacks cyber-criminals have been identified registering large numbers of website domains containing the words `covid' and `coronavirus'\cite{checkpoint2020domain}. Such domains are likely to be believable, and therefore accessed, especially if paired with reputable wording such as WHO or Centers for Disease Control and Prevention (CDC) or key words (e.g., Corona-virusapps.com, anticovid19-pharmacy.com, which have been highlighted as in use\cite{forbes2020highrisk}).  Communications platforms, such as Zoom, Microsoft and Google, have also been impersonated, both through emails and domain names\cite{checkpoint2020domain}. This is noteworthy given the fact that these are the primary technologies used by millions across the world to communicate, both for work and pleasure. These facts, in combination with convincing social engineering emails, text messages and links, provide several notable avenues for criminals to attack. Pharming attacks were much less common but did occur in 13\% of cases. As can be seen Table~\ref{tableOfAttacks}, these often occur alongside other attacks.

COVID-19-inspired fraud has leveraged governmental/scientific announcements to exploit the anxieties of users and seek financial benefit. From our analysis, fraud was typically committed through phishing and email attacks---we also can see this in our sequencing above. In one case, criminals posed as the CDC in an email and politely requested donations to develop a vaccine, and also that any payments be made in Bitcoin\cite{TableJoeBBC2020}. Typical phishing techniques were used, but on this occasion these included requests for money: ``\textit{Funding of the above project is quite a huge cost and we plead for your good will donation, nothing is too small}''. A notable point about this particular attack is that it also ask recipients to share the message with as many people as possible. This is concerning given that people are more likely to trust emails they believe have been vetted by close ones\cite{nursecybercrime2019}.

There were a range of other fraud attempts, largely based on threats or appeals.  For instance, our analysis identified offers of investment in companies claiming to prevent, detect or cure COVID-19, and investment in schemes/trading options which enable users to take advantage of a possible COVID-19 driven economic downturn \cite{USDOJ2020}. There were offers of cures, vaccines, and advice on effective treatments for the virus. The Food and Drugs Administration (FDA) issued 16 warning letters between 6\textsuperscript{th} March and 1\textsuperscript{st} April 2020 to companies \textit{``for selling fraudulent products with claims to prevent, treat, mitigate, diagnose or cure''} COVID-19\cite{2016study,FDA2020}. The European Anti-Fraud Office (OLAF) has responded to the flood of fake products online by opening an enquiry concerning imports of fake products due to COVID-19 pandemic\cite{olaf2020fakeps}, and in the UK, the Medical and Healthcare products Regulatory Agency (MHRA) has began investigating bogus or unlicensed medical devices currently being traded through unauthorised and unregulated websites\cite{ukgov2020fakemed}. 

Extortion attacks were witnessed in our analysis but were less prevalent (appearing in only 13\% of cases) compared to the others above. The most prominent case of this attack was an extortion email threatening to infect the recipient and their family members with COVID-19 unless a Bitcoin payment is made\cite{Sophos2020}. To increase the believability of the message, it included the name of the individual and one of their passwords  (likely gathered from a previous password breach). After demanding money, the message goes on to state: ``\textit{If I do not get the payment, I will infect every member of your family with coronavirus}''. This attempts to use fear to motivate individuals to pay, and uses passwords (i.e., items that are personal) to build confidence in the criminal's message. 

Malware related to COVID-19 increased in prominence during the pandemic and impacted individuals and organisations across the world. As shown above, it was the second largest cyber-attack type, appearing in 65\% of cases. \textit{Vicious panda} and \textit{MBR Loader} were the only new malware discovered in this period. The remaining malware attacks were variants of existing malware and included \textit{Metaljack, REMCOS, Emotet, LOKIBOT, CXK-NMSL, Dharma-Crysis, Netwalker, Mespinoza/Pysa, SpyMax} (disguised as the \textit{Corona live 1.1} app) \textit{GuLoader, Hawkeye, FORMBOOK, Trickbot} and \textit{Ginp}. Ransomware, in particular, was a notable threat and an example of such was COVIDLock, an Android app disguised as a heat map which acted as ransomware; essentially locking the user's screen unless a ransom was paid\cite{Saleh2020}. 

At the organisational level, ransomware has significantly impacted healthcare services---arguably the most fragile component of a country's critical national infrastructure at this time. Attacks have been reported in the United States, France, Spain and the Czech Republic\cite{incisivem2020spa,wired2020hackp}, and using ransomware such as \textit{Netwalker}. Such attacks fit a criminal modus operandi if we assume that malicious actors will target areas where they believe they stand to capitalise on their attacks; i.e., health organisations may be more likely to pay ransoms to avoid loss of patient lives. Interestingly there have since been promises from leading cyber-crime gangs that they will not (or stop) targeting healthcare services. In one report, operators behind CLOP Ransomware, DoppelPaymer Ransomware, Maze Ransomware and Nefilim Ransomware stressed that they did not (normally) target hospitals, or that they would pause all activity against healthcare services until the virus stabilises\cite{bleepingcom2020gangs}.  

Other notable malware examples during the pandemic include: \textit{Trickbot}, a trojan that is typically used as a platform to install other malware on victims' devices---according to Microsoft, \textit{Trickbot} is the most prolific malware operation that makes use of COVID-19 themed lures for its attacks\cite{infosec2020trick}; a Master Boot Record (MBR) rewriter malware that wipes a device's disks and overwrites the MBR to make them no longer usable\cite{SonicWall2020}; and \textit{Corona Live 1.1}, an app that leveraged a legitimate COVID-19 tracker released by John Hopkins University and accessed device photos, videos, location data and the camera\cite{Ng2020}. As the pandemic continues, there are likely to be more strains of malware, targeting various types of harm, e.g., physical, financial, psychological, reputational (for businesses) and societal\cite{agrafiotis2018taxonomy}.

During the COVID-19 pandemic our analysis only identified a very small amount (5\%) of DoS attacks, but there were several reports of hacking. These reports suggested that hacking was not indiscriminate but instead, targeted towards institutions involved in research on coronavirus.

In one report, FBI Deputy Assistant Director stated, ``\textit{We certainly have seen reconnaissance activity, and some intrusions, into some of those institutions, especially those that have publicly identified themselves as working on COVID-related research}''\cite{reuters2020fbi}. This was further supported by a joint security advisory a month later from the UK's NCSC and USA's CISA\cite{ncsccisa2020apt}. In this advisory, Advanced Persistent Threat (APT) groups---some of which may align with nation states---were identified as targeting pharmaceutical companies, medical research organisations, and
universities involved in COVID-19 response. The goal was not necessarily to disrupt their activities (as with the ransomware case), but instead to steal sensitive research data or intellectual property (e.g., on vaccines, treatments). 

While a detailed analysis of these attacks has not yet surfaced, password spraying (a brute-force attack which applying commonly-used passwords in attempting to login to accounts) and exploiting vulnerabilities in Virtual Private Network (VPN) have been flagged\cite{ncsccisa2020apt}. Attribution is another important consideration during such attacks. Determining the true origin of cyber-attacks has always been difficult, however, in response to these COVID-19-related threats, the US openly named the People's Republic of China (PRC) as a perpetrator in a joint FBI/CISA announcement\cite{fbi2020prcip}.

\section{Impact on workforce \label{ImpactOnWorkforce}}

\begin{figure*}[!htb]
    \centering
    \includegraphics[width=0.8\linewidth]{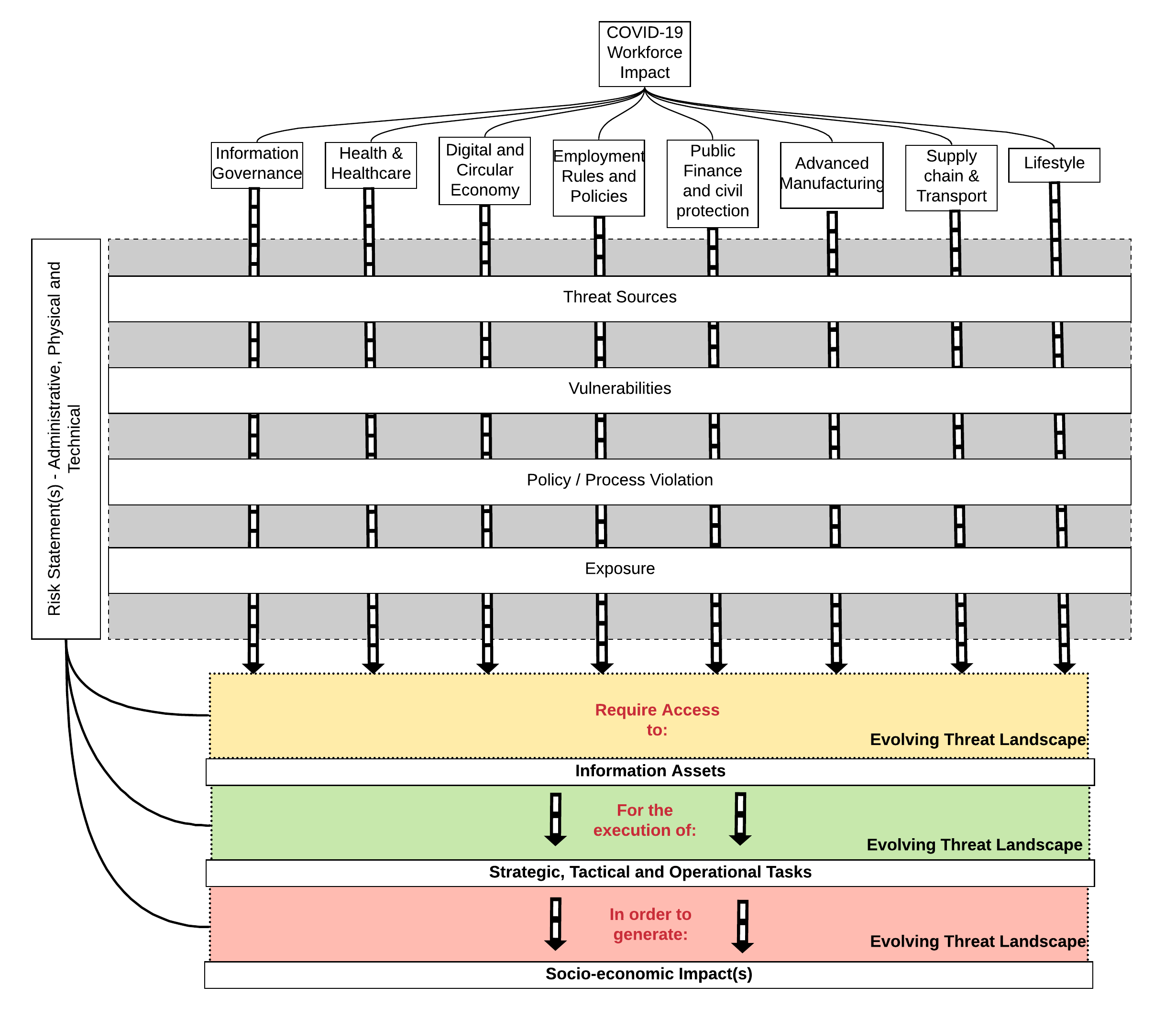}
    \caption{COVID-19 Impact on Workforce}    \label{workforce}
\end{figure*}

The effects of the pandemic, the mass quarantine of staff and the measures put in place to facilitate remote working and resilience of existing cyber-infrastructures, against the attacks and timelines previously described, had a profound effect on the workforce --  the people engaged in or available for work. The pandemic also had an effect on the resilience of technology, socio-economic structures and threatened, to a certain degree, the way people live and communicate. Figure \ref{workforce} illustrates the COVID-19 impact on the workforce across eight different categories. All categories seemingly integrate with cyber-enabled assets and tools and different categories may be impacted differently. The pandemic created risk conflicts, for example, strict compliance with security standards which discourage data sharing, could be more harmful than sharing the data. So, whilst there may be strict requirements for patient data not to be accessed at home by GPs (general practitioners), this causes a greater harm during quarantine than enabling GPs to access patient data. Also, the way confidential patient information is processed requires a data protection impact assessment (DPIA) to enable further NHS support where needed. This can have an impact in terms of the timely delivery of medical interventions in response to COVID-19.

In traditional risk classification, elements like asset registration and valuation, threat frequency and vulnerability probability are at greater risk of cyber threat. We, therefore, anticipate changes on the way the workforce accesses those information assets and how strategic, tactical and operational tasks are executed to generate socio-economical outputs. These changes can be captured by the development and testing of risk statements capturing 1) threat agents, 2) vulnerabilities, 3) Policy/process violation and 4) overall asset exposure on all emerging threat landscapes as illustrated in Figure \ref{workforce}. These changes unavoidably cascade further changes to the threat landscapes associated with remote workforce activities and the increasing frequency of weaponised attack vectors related to the coronavirus spreading. Given the current climate, it is difficult to predict whether these changes will have a long-lasting effect on the workforce, but their significance is already recorded \cite{wge}. Therefore, it is increasingly important that the control of information (storage, processing, transmission) has an elevated importance given the increase of cyber-attacks on important infrastructures. 

Governments, private and public sectors throughout Europe currently consider measures to contain and mitigate COVID-19 impact on existing data structures and information governance frameworks (for example, \cite{DataProtection2020}). Particular emphasis is placed upon the implications of the pandemic in the processing of personal data. The General Data Protection Regulation (GDPR) legislation in the UK dictates that personal data must be processed only for the specific and explicit purposes for which it has been obtained \cite{ICO2020}. In addition, data subjects should always receive explicit and transparent information with regard to the processing activities undertaken, including that of features and nature of the activity, retention period and purpose of processing. There are challenges related to the governance legal and regulatory compliance landscape in terms of conformance versus rapid access and processing of data by different entities. This is quite apparent in cases where public authorities seek to obtain PII to reduce the spread of COVID-19. Typical examples also include contact tracing applications and platforms in which the data is aggregated online for post-processing \cite{Andrea2020}. Specific legislative measures have to be re-deployed or introduced to safeguard public security while maintaining privacy at scale, while legal and regulatory principles continue to upheld \cite{cvfd}. 

With the rapid increase of COVID-19 symptoms, governments had to derive a plan that would enable them to understand epidemiological data further and identify positive interventions to contain and mitigate the impact of the pandemic. Research shows a high correlation between the use of big data that includes private identifiable information in the effectiveness of these epidemiological investigations \cite{gred}. That meant that in most cases, citizens had to provide this information voluntarily and that quickly resulted in discussions and debates on the trade-offs between public safety versus personal privacy \cite{ahn2020balancing}. The information has also been obtained through internet communication technology. Medical testing equipment and coronavirus testing at a large-scale were used as instruments for data collection in the fight to reduce mortality rates. The legal and regulatory compliance frameworks differ between countries; thus, managing personal information was subject to different privacy protection measures. 

The de-identification of personal information was another component that governments had to exercise to satisfy personal privacy requirements and increase the trust of human participants during the epidemiological investigations. The process of collecting and processing personal information by applying de-identification technologies raised technical challenges with regards to accuracy and consent, safe and legally defensive data disposal and robustness of associated policies of data processing and management for epidemiological research. The urgency of the situation and the speed at which the data had to be acquired and processed, created a sense of distrust amongst citizens and challenged the efficacy of the existing processes in place \cite{ahn2020balancing}. The extensive lockdown periods introduced in many countries (described in Section \ref{sec:timeline}) have also tested their ability to deploy strategies for business recovery after these periods. These strategies had to ensure smooth and phased out recovery within an ongoing pandemic, which has proved to be a challenging task. However, there is an unprecedented speed and scale on the R\&D activities in response to the COVID-19 outbreak forcing cross-organizational multilateral collaborations  \cite{AstraZeneca2020, Kituyi2020}. 

There is currently a challenge across Europe to orchestrate information sharing in a timely and accurate manner as even mainstream media sources seem to have propagated false information \cite{svd}. The increase on both frequency and impact of these attacks will test further our existing monitoring and auditing capabilities, logical and physical access controls, authentication and verification schemes currently deployed. Also, as part of the current enterprise risk management approaches the way organisations sanitise incident reporting, media disposal and data destruction and sharing processes will also be tested alongside to traditional defence-in-depth principles currently established as de-facto. The finance sector is also affected as the predicted financial recession will leverage the sophistication and scale of targeted attacks as threat actors grow their capabilities \cite{sdbv}. 

\section{Conclusion and Future work \label{sec:conclusion}}

The COVID-19 pandemic has generated remarkable and unique societal and economic circumstances leveraged by cyber-criminals. Our analysis of events such as announcements and media stories has shown what appears to be a loose correlation between the announcement and a corresponding cyber-attack campaign which utilises the event as a hook thereby increasing the likelihood of success. 

The COVID-19 pandemic, and the increased rate of cyber-attacks it has invoked have wider implications, which stretch beyond the targets of such attacks.  Changes to working practises and socialization, mean people are now spending increased periods of time online.  In addition to this, rates of unemployment have also increased, meaning more people are sitting at home online- it is likely that some of these people will turn to cyber-crime to support themselves.  The combination of increased levels of cyber-attacks and cyber-crime means there may be implications for policing around the World- law enforcement must ensure it has the capacity to deal with cyber-crime \cite{sipr2020}.

The analysis presented in this paper has highlighted a common modus-operandi of many cyber-attacks during this period. Many cyber-attacks begin with a phishing campaign which directs victims to download a file or access a URL. The file or the URL act as the carrier of malware which, when installed, acts as the vehicle for financial fraud. The analysis has also shown that to increase the likelihood of success, the phishing campaign leverages media and governmental announcements. 

Although this analysis is not necessarily novel, we believe this is the first time that this has been supported with a context of actual live events. This analysis gives rise to the recommendation that governments, the media and other institutions  should be aware that announcements and the publication of stories are likely to give rise to the perpetration of associated cyber-attack campaigns which leverage these events. The events should be accompanied by a note / disclaimer outlining how information relating to the announcement will be relayed.

Our research presents opportunity for further research. This research has shown what can best be described as a loose direct and inverse correlation between events and cyber-attacks. Further research should investigate this phenomenon and outline whether a predictive model can be used to confirm this relationship. There is an abundant supply of cyber-attack case studies relating to countries around the world and a wider analysis of the problem can help in affirming this phenomenon.

	\small{ 
	    \bibliographystyle{IEEEtran}
		\bibliography{references}}
\end{document}